# SnAs-based layered superconductor NaSn$_2$As$_2$

Yosuke Goto, Akira Yamada, Tatsuma D. Matsuda, Yuji Aoki, and Yoshikazu Mizuguchi*

Department of Physics, Tokyo Metropolitan University, Hachioji 192-0397, Japan

*E-mail: mizugu@tmu.ac.jp

**Abstract**

  Superconductivity with exotic properties has often been discovered in materials with a layered (two-dimensional) crystal structure. The low dimensionality affects the electronic structure of materials, which could realize a high transition temperature ($T_c$) and/or unconventional pairing mechanisms. Here, we report the superconductivity in a layered tin arsenide NaSn$_2$As$_2$. The crystal structure consists of (Sn$_2$As$_2$)$^{2-}$ bilayers, which is bound by van-der-Waals forces, separated by Na$^+$ ions. Measurements of electrical resistivity and specific heat confirm the bulk nature of superconductivity of NaSn$_2$As$_2$ with $T_c$ of 1.3 K. Our results propose that the SnAs layers will be a basic structure providing another universality class of a layered superconducting family, and it provides a new platform for the physics and chemistry of low-dimensional superconductors with lone pair electrons.

A layered crystal structure is an attractive stage to explore superconductors with a high transition temperature ($T_c$) and to discuss the mechanisms of unconventional superconductivity, as exemplified by the cuprates [1] and the Fe-based superconductors [2]. The discovery of a basic structure, which works as a superconducting layer, such as the $CuO_2$ plane and the $Fe_2An_2$ (An = P, As, S, Se, Te) layer, have opened new physics and chemistry fields on low-dimensional superconductors because many structural analogous could be designed by changing the structure or the alignment of the spacer layers as well as superconducting layers. Since 2012, our group has developed $BiCh_2$-based (Ch = S, Se) layered superconductors [3-5]. Although early studies have suggested conventional pairing mechanisms of the $BiCh_2$-based compounds [6-10], recent studies including the first-principles calculations [11], angle-resolved photoemission spectroscopy [12], and the isotope effect [13] have indicated unconventional pairing mechanisms in the $BiCh_2$-based superconductors. One of the possible mechanisms is charge-fluctuation-mediated one, which was proposed from neutron studies of in-plane local structure of $RE(O,F)BiS_2$ [14-16], where RE represent rare earth element. The locally distorted in-plane (Bi-Ch plane) structure should be originated from the structural instability due to $Bi^{3+}$ lone pair electrons. The studies on chemical pressure effects in $RE(O,F)BiS_2$ suggested that the activation/deactivation of lone-pair effects would be related to the emergence of superconductivity [17]. In addition, the presence of lone-pairs makes a van-der-Waals (vdW) gap between the BiCh planes; hence the crystal of the $BiCh_2$-based superconductors can be readily exfoliated, which is useful to investigate electronic properties with surface sensitive probes [12,18] and develop other functionalities.

In this letter, we present a novel layered superconductor $NaSn_2As_2$ containing $Sn^{2+}$ ions with lone-pair electrons. $NaSn_2As_2$ crystallizes in a trigonal $R\bar{3}m$ unit cell characterized by SnAs bilayers separated by six-coordinated $Na^+$ cations, as schematically shown in the inset of Figure 1 [19,20]. The SnAs bilayers are bound by vdW forces with an approximately 3.3 Å gap between the Sn atoms of the adjacent layers. Assuming a +2 oxidation state for Sn, and a −3 oxidation state for As, conducting SnAs layer is negatively charged as $[Sn_2As_2]^{2-}$. Thus, $NaSn_2As_2$ is not electron-balanced compounds, as in the case of recently reported $Li_{1-x}Sn_{2+x}As_2$ [21], which is in contrast to the isostructural electron-balanced compounds $SrSn_2As_2$ [22] or $EuSn_2As_2$ [23]. Because of the vdW gap between SnAs layers, these compounds can be readily exfoliated through both mechanical and liquid-phase methods [20,23]. Density functional theory calculation indicated that Fermi surface of these compounds dominantly consists of hybridization between Sn and As orbitals [20,23]. Because there exists various structural analogous compounds with conducting SnPn layers (Pn = P, As, Sb), such as ASn$_2$Pn$_2$ (A = Li, Na, K, Sr, Eu [19–25]), ASnPn [22,26,27], and Sn$_4$Pn$_3$ [28,29], the discovery of superconductivity in NaSn$_2$As$_2$ will be a starting point of new physics and chemistry of the layered superconductor family.

Single crystals of NaSn$_2$As$_2$ were synthesized by melting raw elements. Surface oxide layer of Na (Sigma-Aldrich, 99.9%) was mechanically cleaved before use. Sn (99.99%) and As (99.9999%) were purchased from Kojundo Chemical and used without further purification. The raw elements were handled in Ar-filled glovebox with a gas-purifier system. Stoichiometric ratio of Na, Sn, and As was sealed in an evacuated quartz tube, heated at 750 °C for 20 h, and cooled to 350 °C in 10 h. Energy dispersive X-ray spectroscopy shows the chemical composition of obtained products was Na : Sn : As = 1 : 1.89(8) : 2.12(7). Phase purity of the sample was examined using powder X-ray diffraction (PXRD; Rigaku miniflex 600) using powders prepared by grinding the obtained crystals. Lattice parameters were calculated using Rietveld refinement using RIETAN-FP [30]. Single-crystal XRD (Rigaku XtaLAB) for a selected small single crystal with dimensions of 0.06 × 0.1 × 0.07 mm was analyzed using the program SHELX-97 [31]. Temperature ($T$) dependence of electrical resistivity ($\rho$) of sample flake with dimensions of 3 × 1 × 0.04 mm was measured using four-terminal method with a physical property measurement system (PPMS; Quantum Design) equipped with a $^3$He-probe system. Magnetic field was applied along the $c$-axis direction of the crystal. Specific heat ($C$) versus $T$ was measured by a relaxation method with PPMS.

Figure 1 shows PXRD pattern of the obtained sample. Almost all the diffraction peaks are assigned to those of the NaSn$_2$As$_2$ phase, indicating that the obtained sample is nearly single-phase. However, tiny diffraction peaks attributable to elemental Sn were also observed, as indicated by the asterisk in the Fig. 1. Lattice parameters calculated by the Rietveld refinement of the PXRD pattern were $a$ = 4.00409(10) Å and $c$ = 27.5944(5) Å. Structural parameters obtained from single-crystal XRD are summarized in Table 1. Basically, these structural parameters are in good agreement with the previous reports [19,20]. We cannot exclude the possibility of several percent of Na deficiency in the present structural refinement. We notice that, however, refinements with partial substitution of the Sn site by Na did not improve the structural refinement, unlike the case of isostructural Li$_{1–x}$Sn$_{2+x}$As$_2$ [21].

Figure 2(a) and (b) show $\rho$–$T$ plots of NaSn$_2$As$_2$. Metallic behavior in the electrical resistivity was observed at temperatures >10 K. A sharp drop in $\rho$ was seen at 1.3 K, accompanied by zero resistivity below 1.1 K, indicating a superconducting transition. Note that steep drop except for this temperature range was not observed (Figure 2(b)), although small amount of elemental Sn with $T_c$ of 3.7 K was detected using PXRD. The transition temperature of NaSn$_2$As$_2$ shifts to lower temperatures with increasing magnetic field, as shown in Figure 2(c). $T_c^{90\%}$ and $T_c^{zero}$ obtained from the temperature dependences of electrical resistivity under magnetic fields up to 3000 Oe are presented in Figure 2(d). Here, $T_c^{90\%}$ is defined as the temperature where $\rho$ becomes 90% of that at 3 K (normal states), as indicated with a dashed line in Fig. 3(c). $T$ dependence of upper critical field ($H_{c2}$) is still almost linear at $T \approx 0.5$ K. From the Werthamer Helfand Hohemberg (WHH) model [33], the $H_{c2}$ at 0 K was estimated to be $H_{c2}(0) \sim 2000$ Oe. From the $H_{c2}$, the coherent length was estimated to be ~41 nm using the equation of $\xi^2 = \Phi_0/2\pi\mu_0 H_{c2}$. Note that a kink was observed in $\rho$–$T$ data under 1500 and 2000 Oe at around 0.7 K (Fig. 2(c)). This is probably because of highly anisotropic superconducting states due to the layered structure; $H_{c2}$ is believed to be higher when magnetic field is applied parallel to *ab*-plane. Although magnetic field was applied along the *c*-axis of the crystal, some parts of domain may tilt with respect to magnetic field. Another possibility is local inhomogeneity in the nanoscale, which is revealed for Li$_{1-x}$Sn$_{2+x}$As$_2$ by combinational analysis using nuclear magnetic resonance, transmission electron microscopy, and neutron diffraction [21]. The $\rho$–$T$ measurements under magnetic field parallel to the *ab* plane were not performed in the present study because of the limitation of measurement apparatus.

Figure 3(a) shows the $C/T$ as a function of $T$ under magnetic fields of 0 and 30000 Oe. A steep jump of $C/T$ is observed at $T_c = 1.3$ K, which is consistent with the superconducting transition seen in $\rho$–$T$ data. Figure 3(b) shows the $C/T$–$T^2$ plot measured at 30000 Oe. By performing least-squares fitting to $C/T = \gamma + \beta T^2$, where $\gamma$ and $\beta$ represent the electronic specific heat parameter and the phonon contribution parameter, respectively, we obtained those parameters as $\gamma = 3.97$ mJmol$^{-1}$K$^{-2}$ and $\beta = 1.120$ mJmol$^{-1}$K$^{-4}$. Debye temperature $\Theta_D = (12\pi^4 r N_A k_B/5\beta)^{1/3}$ was estimated to be 205 K, where $r = 5$ is the number of atom per formula unit, $N_A$ is the Avogadro constant, and $k_B$ is the Boltzmann constant, respectively. $C_{el}$, the electron contribution to specific heat, is determined by subtracting the phonon contribution, which can be described as $\beta T^3$ when $T \ll \Theta_D$, from total $C$. Figure 3(c) shows the $C_{el}/T$–$T$ plot. The specific heat jump at $T_c$ ($\Delta C_{el}$) is 7.82 mJmol$^{-1}$K$^{-2}$. From the obtained parameters, $\Delta C_{el}/\gamma T_c$ is calculated as 1.50, which is slightly larger than the value expected from the weak-coupling BCS approximation ($\Delta C_{el}/\gamma T_c = 1.43$). The slightly larger superconducting gap may result from the strong-coupling nature of superconductivity, but it seems reasonable as a fully gapped superconductor. The

electron–phonon coupling constant ($\lambda$) can be determined by Macmillan's theory [34], which gives

$$\lambda = \frac{1.04 + \mu^* \ln(\Theta_D/1.45T_c)}{(1 - 0.62\mu^*)\ln(\Theta_D/1.45T_c) - 1.04}$$

where $\mu^*$ is defined as the Coulomb pseudopotential. Taking $\mu^* = 0.13$ gives $\lambda = 0.44$, which is consistent with weak or intermediate coupling strength of BCS superconductivity. As a result, a bulk nature of superconductivity at $T_c = 1.3$ K in $NaSn_2As_2$ has been confirmed from electrical resistivity and specific heat measurements.

We briefly discuss the difference in crystal structure between $NaSn_2As_2$ and $Li_{1-x}Sn_{2+x}As_2$ by comparing the experiences on layered FeAs-based compounds LiFeAs [35] and NaFeAs [36]. In the LiFeAs phase, Li deficiency is present and the obtained sample ($Li_{1-x}FeAs$) shows superconductivity without any magnetic ordering [35], which indicates that Li deficiency results in carrier doping and suppresses antiferromagnetic ordering, which is a typical feature of a parent phase of the FeAs-based superconductor. In contrast, NaFeAs is almost stoichiometric and exhibits antiferromagnetic ordering, and Co-doped $NaFe_{2-x}Co_xAs_2$ becomes superconductive [36]. The different physical properties may result from the different ionic radii: 92 pm and 118 pm for $Li^+$ and $Na^+$ when coordination number is 8. Based on these differences in alkaline site deficiency in Li- and Na-containing FeAs systems, we speculate that our $NaSn_2As_2$ sample is almost stoichiometric, similarly to NaFeAs, and is almost free of site deficiency or solution with Sn, while Li/Sn solution was reported in $Li_{1-x}Sn_{2+x}As_2$ [21].

Here, we have demonstrated the discovery of the novel layered superconductor $NaSn_2As_2$. Because there exist various structural analogues containing SnPn layeres, superconductivity in the SnPn layers may generally emerge when carriers are doped in those SnAs-based layered compounds. Furthermore, three-dimensional Dirac semimetal state and non-tribial topological phase have been predicted in these compounds [37,38]. Our discovery presented here will open up a new platform of the physics and chemistry of low-dimensional superconductors and related notable phenomena.

**Acknowledgement**
This work was partly supported by Grants-in-Aid for Scientific Research (Nos. 15H05886, 15H05884, 16H04493, 17K19058, 16K17944, and 15H03693), JST-PRESTO (No. JPMJPR16R1), and JST-CREST (No. JPMJCR16Q6), Japan.

**References**

1) J. B. Bednorz and K. Müller, Z. Phys. B **64**, 189 (1986).
2) Y. Kamihara, T. Watanabe, M. Hirano, and H. Hosono, J. Am. Chem. Soc. **130**, 3296 (2008).
3) Y. Mizuguchi, H. Fujihisa, Y. Gotoh, K. Suzuki, H. Usui, K. Kuroki, S. Demura, Y. Takano, H. Izawa, and O. Miura, Phys. Rev. B **86**, 220510 (2012).
4) Y. Mizuguchi, S. Demura, K. Deguchi, Y. Takano, H. Fujihisa, Y. Gotoh, H. Izawa, and O. Miura, J. Phys. Soc. Jpn. **81**, 114725 (2012).
5) Y. Mizuguchi, J. Phys. Chem. Solids **84**, 34 (2015).
6) H. Usui and K. Kuroki, Nov. Supercond. Mater. **1**, 50 (2015).
7) X. Wan, H. C. Ding, S. Savrasov, and C. G. Duan, Phys. Rev. B **87**, 115124 (2013).
8) S. F. Wu, P. Richard, X. B. Wang, C. S. Lian, S. M. Nie, J. T. Wang, N. L. Wang, and H. Ding, Phys. Rev. B **90**, 54519 (2014).
9) G. Lamura, T. Shiroka, P. Bonf, S. Sanna, R. De Renzi, C. Baines, H. Luetkens, J. Kajitani, Y. Mizuguchi, O. Miura, K. Deguchi, S. Demura, Y. Takano, and M. Putti, Phys. Rev. B **88**, 180509 (2013).
10) T. Yamashita, Y. Tokiwa, D. Terazawa, M. Nagao, S. Watauchi, I. Tanaka, T. Terashima, and Y. Matsuda, J. Phys. Soc. Japan **85**, 73707 (2016).
11) C. Morice, R. Akashi, T. Koretsune, S. S. Saxena, and R. Arita, Phys. Rev. B **95**, 180505 (2017).
12) Y. Ota, K. Okazaki, H. Q. Yamamoto, T. Yamamoto, S. Watanabe, C. Chen, M. Nagao, S. Watauchi, I. Tanaka, Y. Takano, and S. Shin, Phys. Rev. Lett. **118**, 167002 (2017).
13) K. Hoshi, Y. Goto, and Y. Mizuguchi, arXiv: 1708.08252.
14) A. Athauda, J. Yang, S. Lee, Y. Mizuguchi, K. Deguchi, Y. Takano, O. Miura, D. Louca, Phys. Rev. B **91**, 144112 (2014).
15) A. Athauda, C. Hoffman, Y. Ren, S. Aswartham, J. Terzic, G. Cao, X. Zhu, D. Louca, J. Phys. Soc. Jpn. **86**, 054701 (2017).
16) A. Athauda, Y. Mizuguchi, M. Nagao, J. Neuefeind, D. Louca, arXiv:1708.03957
17) Y. Mizuguchi, E. Paris, T. Sugimoto, A. Iadecola, J. Kajitani, O. Miura, T. Mizokawa, and N. L. Saini, Phys. Chem. Chem. Phys. **17**, 22090 (2015).
18) T. Machida, Y. Fujisawa, M. Nagao, S. Demura, K. Deguchi, Y. Mizuguchi, Y. Takano, H. Sakata, J. Phys. Soc. Jpn. **83**, 113701 (2014)
19) M. Asbrand, B. Eisenmann, and J. Klein, Z. Anorg. Allg. Chem. **621**, 576 (1995).
20) M. Q. Arguilla, J. Katoch, K. Krymowski, N. D. Cultrara, J. Xu, X. Xi, A. Hanks, S. Jiang, R. D. Ross, R. J. Koch, S. Ulstrup, A. Bostwick, C. Jozwiak, D. W. Mccomb, E. Rotenberg, J. Shan, W. Windl, R. K. Kawakami, and J. E. Goldberger, ACS Nano

**10**, 9500 (2016).

21) K. Lee, D. Kaseman, S. Sen, I. Hung, Z. Gan, B. Gerke, R. Po, M. Feygenson, J. Neuefeind, O. I. Lebedev, and K. Kovnir, J. Am. Chem. Soc. **137**, 3622 (2015).

22) B. Eisenmann and J. Klein, Z. Anorg. Allg. Chem. **598/599**, 93 (1991).

23) M. Q. Arguilla, N. D. Cultrara, Z. J. Baum, S. Jiang, R. D. Ross, and J. E. Goldberger, Inorg. Chem. Front. **2**, 378 (2017).

24) P. C. Schmidt, D. Stahl, B. Eisenmann, R. Kniep, V. Eyert, and J. Kübler, J. Solid State Chem. **97**, 93 (1992).

25) M. Asbrand, F. J. Berry, B. Eisenmann, R. Kniep, L. E. Smart, and R. C. Thied, J. Solid State Chem. **118**, 397 (1995).

26) Z. Lin, G. Wang, C. Le, H. Zhao, N. Liu, J. Hu, L. Guo, and X. Chen, Phys. Rev. B **95**, 165201 (2017).

27) K. H. Lii, R. C. Haushalter, J. Solid State Chem. **67**, 374 (1987).

28) K. Kovnir, Y. V. Kolen'ko, A. I. Baranov, I. S. Neira, A. V. Sobolev, M. Yoshimura, I. A. Presniakov, and A. V. Shevelkov, J. Solid State Chem. **182**, 630 (2009).

29) O. Olofsson, Acta Chem. Scand. **24**, 1153 (1970).

30) F. Izumi and K. Momma, Solid State Phenom. **130**, 15 (2007).

31) G. M. Sheldrick: SHELX-97: Program for the Solution for Crystal Structures (University of Gottingen, Germany, 1997).

32) K. Momma and F. Izumi, J. Appl. Crystallogr. **41**, 653 (2008).

33) N. R. Werthamer, E. Helfand, and P. C. Hohemberg, Phys. Rev. **147**, 295 (1966).

34) W. L. McMillan, Phys. Rev. **167**, 331 (1967).

35) X. C. Wang, Q. Q. Liu, Y. X. Lv, W. B. Gao, L. X. Yang, R. C. Yu, F. Y. Li, and C. Q. Jin, Solid State Commun. **148**, 538 (2008).

36) D. R. Parker, M. J. P. Smith, T. Lancaster, A. J. Steele, I. Franke, P. J. Baker, F. L. Pratt, M. J. Pitcher, S. J. Blundell, and S. J. Clarke, Phys. Rev. Lett. **104**, 57007 (2010).

37) Q. D. Gibson, L. M. Schoop, L. Muechler, L. S. Xie, M. Hirschberger, N. P. Ong, R. Car, and R. J. Cava, Phys. Rev. B **91**, 205128 (2015).

38) X. Dai, C. Le, X. Wu, S. Qin, Z. Lin, and J. Hu, Chinese Phys. Lett. **33**, 127301 (2016).

**Table captions**

Table 1

Structural parameters of NaSn$_2$As$_2$ at room temperature determined by single-crystal XRD measurements. $R_1$ and $wR_2$ are reliability factors and $B$ is the equivalent isotropic atomic displacement parameter. Standard deviations are given in parentheses. Site occupancy of each site is unity.

**Figure captions**

Figure 1

PXRD pattern of NaSn$_2$As$_2$. Vertical tick marks represent the Bragg diffraction angles of NaSn$_2$As$_2$. Asterisk denote the diffraction peak due to elemental Sn. Inset shows the schematic representation of crystal structure of NaSn$_2$As$_2$ ($R\bar{3}m$ space group). The black solid line represents the unit cell in the hexagonal setting. Crystal structure was depicted using VESTA [32].

Figure 2

(a) Temperature ($T$) dependence of electrical resistivity ($\rho$) of NaSn$_2$As$_2$. (b) $\rho$–$T$ data below 4 K. (c) $\rho$–$T$ data under magnetic fields ($H$) up to 3000 Oe. Magnetic field was applied along the $c$-axis. Dashed line represents 90% of $\rho$ at 3 K (d) Magnetic field-temperature phase diagram of NaSn$_2$As$_2$. Dashed lines represent the least-squares fits of data plots.

Figure 3

(a) Temperature ($T$) dependence of specific heat ($C$) divided by $T$ for NaSn$_2$As$_2$ at magnetic fields ($H$) of 0 and 30000 Oe. (b) $C/T$–$T^2$ plot measured at 30000 Oe. Red solid line represents the least-squares fitting results. (c) $C_{el}$–$T$ plot measured at 0 Oe. Black solid line is used to estimate the specific heat jump ($\Delta C_{el}$) at $T_c$.

Table 1

| Space group | Lattice parameters | Atom | Site | Symmetry | x | y | z | $B$ (Å$^2$) |
|---|---|---|---|---|---|---|---|---|
| Trigonal $R\bar{3}m$ (No. 166) | $a$ = 4.010(4) Å | Na | 3a | $-3m$ | 0 | 0 | 0 | 1.5(3) |
| Hexagonal setting | $c$ = 27.56(4) Å | Sn | 6c | $3m$ | 0 | 0 | 0.21021(6) | 1.69(13) |
| $z$ = 3 | $\gamma$ = 120° | As | 6c | $3m$ | 0 | 0 | 0.40742(13) | 1.58(14) |
| $R_1$ = 7.67% | $wR_2$ = 27.9% | | | | | | | |

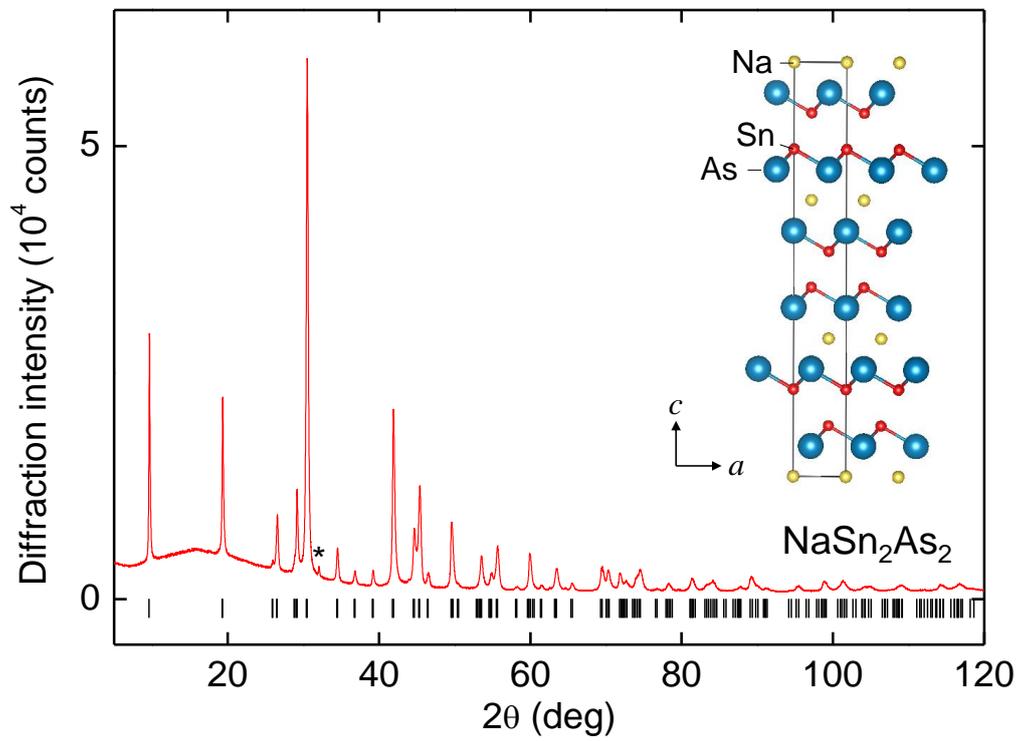

Figure 1

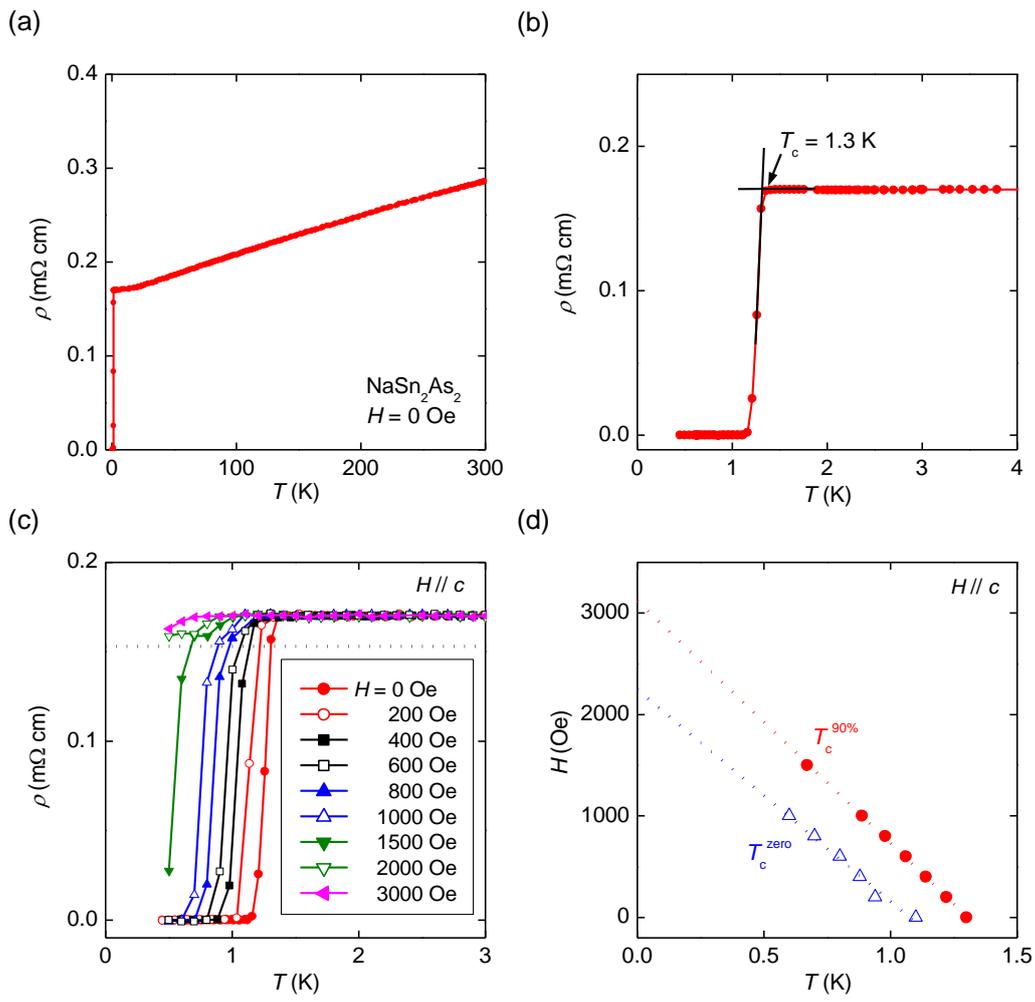

Figure 2

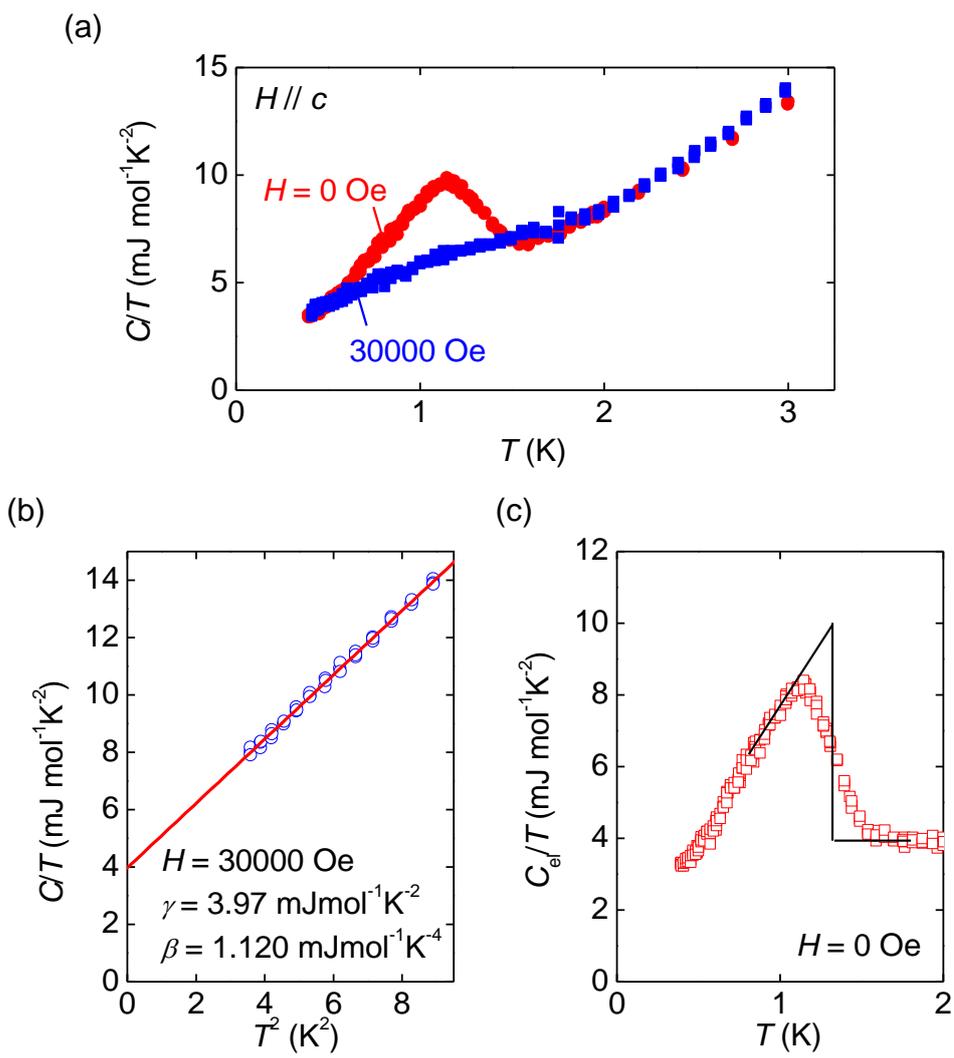

Figure 3